\documentclass[aps,pra,reprint,twocolumn,showpacs,showkeys,floatfix,nobibnotes]{revtex4-2}
\usepackage[utf8]{inputenc}
\usepackage{amsfonts}
\usepackage{multirow}
\usepackage{float}
\usepackage{bm}
\usepackage{amssymb,amsmath}
\usepackage{graphicx,mathptmx,physics}
\usepackage[
colorlinks=true,
citecolor=blue,
linkcolor=blue,
urlcolor=blue
]{hyperref}
\usepackage{xcolor}
\usepackage{etoolbox}
\allowdisplaybreaks 
\begin{document}
	
	\title{A phase-space approach for performing continuous-variable quantum teleportation with a non-Gaussian resource}

\author{Ankita}
\author{Arpita Chatterjee}
\email{Corresponding author: arpita.sps@gmail.com}

\affiliation{Department of Mathematics, J. C. Bose University of Science and Technology, YMCA, Faridabad 121006, Haryana, India}
\begin{abstract}
	We present a detailed phase-space analysis of continuous-variable quantum teleportation using a photon-subtracted two-mode squeezed Fock state (PS-TMSFS) as an entangled resource. We investigate the performance of PS-TMSFS within the Braunstein-Kimble teleportation protocol. The resource-state preparation is described and the success probability related to the photon-subtraction process is derived analytically. The Wigner characteristic function of PS-TMSFS is calculated and then employed to determine the fidelity for input Gaussian and non-Gaussian states. The dependence of the success probability and the teleportation fidelity on the squeezing parameter and the beam-splitter transmissivity is analyzed for both symmetric and asymmetric photon-subtraction scenarios. This results that the symmetric photon subtraction consistently outperforms the corresponding asymmetric configurations with the $(2,2)$ configuration providing the highest teleportation fidelity among the considered PS-TMSFS resources. A significant improvement over the Gaussian two-mode squeezed vacuum (TMSV) resource is observed for teleporting Gaussian input states, particularly in the low-squeezing regime, whereas the relative advantage for non-Gaussian input states depends on the functioning resource parameters and gradually diminishes as squeezing increases.

\end{abstract}
\maketitle

\section{Introduction}

Quantum entanglement is a fundamental resource in quantum information science, enabling information-processing tasks beyond the potential of classical systems. In continuous-variable (CV) optical systems, entangled states of light play a crucial role in a variety of quantum communication protocols, including quantum teleportation \cite{Hou2016,Wang2015,Wang2014}, quantum dense coding \cite{Zhang2002,Yeo2006}, and quantum computing \cite{Zhu2005,Tan2017}. Multiple techniquess have been proposed for the preparation of such states \cite{Hu2017, Lin2016, Su2014}.

Among various quantum communication protocols, quantum teleportation (QT) has received significant attention from both the theoreticians and experimentalists \cite{Hu2016}. In CV implementations of QT, the two-mode squeezed vacuum (TMSV) state \cite{Milburn1999} has established itself as one of the most widely used resource states, owing to its deterministic generation through non-linear optical processes \cite{Bowen2003,Eckstein2011}. TMSV belongs to the class of Gaussian states, whose properties are completely characterized by the first and second-order moments of the quadrature operators or equivalently by the covariance matrix in phase space.

Although Gaussian states are accessible both experimentally and theoretically, they impose multiple challenges on the performance of quantum information protocols. In particular, Gaussian states suffer from several limitations including the difficulty of entanglement distillation using Gaussian operations \cite{Eisert2002}, restricted degrees of entanglement due to finite squeezing \cite{Braunstein2005}, and the limited range of non-classical features arising from their description by only first and second-order moments \cite{Weedbrook2012}.

One possible approach to overcome these drawbacks is to move beyond Gaussian states and explore non-Gaussian resources \cite{Manali2025} manufactured by conditional operations such as photon-subtraction \cite{Cochrane2002,Wenger2004,Kitagawa2006,Zhang2010,Park2012,Bartley2013,LeeS2013}, photon-addition \cite{Zhang2011,Zhang2013}, and photon catalysis \cite{Lee2011,LeeJ2013} applied to Gaussian states that can provide stronger non-classical properties \cite{Bartley2015}. Out of these operations, photon-subtraction has attracted major attention due to its relative experimental accessibility and its ability to enhance non-classical features such as entanglement and quantum correlations \cite{Ourjoumtsev2007,Takahashi2010,Kurochkin2014}. Indeed, photon-subtracted states have been shown to improve the performance of different CV quantum information protocols in the context of quantum metrology \cite{Braun2014,Hou2023}, quantum illumination \cite{Zhang2014,Fan2018}, and quantum teleportation \cite{Wang22015}. In many studies, photon subtraction has also been observed to outperform photon addition in enhancing performance metrics \cite{Navarrete2012,Chandan2023}.

So far, most studies have focused on non-Gaussian operations applied to Gaussian states such as TMSV state \cite{Cochrane2002,Dell2007,DellAnno2010,Ankita2026}. It is therefore of interest to consider the entangled states beyond the conventional Gaussian setting. One such possibility is to apply the two-mode squeezing operator to any non-Gaussian excited Fock state ($\ket{n}$ with higher photon-number as $n\geq 1$), resulting in two-mode squeezed Fock states (TMSFS) \cite{KOrolev2024,Bashmakova2025}. These states represent non-Gaussian generalization of the TMSV state and possess an extensive quantum correlations arising from their underlying photon-number structure. Owing to these features, TMSFS and engineered states act as promising quantum channels for CV quantum information processing tasks. Despite their potential advantages, theoretical investigation involving such non-Gaussian states is often computationally challenging. In particular, teleportation fidelity which measures how accurately the original state is reconstructed at the output mode, typically involves complicated operator expressions or multidimensional integrations in phase space. These difficulties become even more pronounced when additional non-Gaussian operations, such as photon-subtraction, are applied to existing non-Gaussian states.

Phase-space method \cite{Ying2019} provides a powerful framework for addressing these challenges. Specifically, the Wigner characteristic function formalism offers a convenient and systematic study of quantum states in phase space \cite{Marian2006,Zubarev2022}. It considerably simplifies the description of the states obtained through non-Gaussian operations. Furthermore, in CV quantum teleportation, the fidelity can be expressed in terms of characteristic functions of the input and resource states, making this approach well suited for analytical investigation. The characteristic function formulation enables one to treat mixed states and multi-mode systems in a unified manner, providing a systematic approach to evaluate the relevant quantities such as correlation functions and teleportation fidelities \cite{Chandan2025}. Therefore, it is worthwhile to employ the Wigner characteristic function to compute the teleportation fidelity associated with PS-TMSFS and to examine its suitability as an entangled resource for quantum communication protocols. 

In this work, we investigate the role of PS-TMSFS as an entangled resource in CV quantum teleportation. We present a technically feasible scheme for generating PS-TMSFS and evaluate its success probability. We then derive the corresponding Wigner characteristic function to compute the teleportation fidelity. To provide a better understanding of the proposed resource, we consider the teleportation of both Gaussian and non-Gaussian input states. While Gaussian state serves as a standard reference for teleportation, non-Gaussian state exhibits superior quantum features and therefore emerges as a more promising initial state to evaluate the resource's performance. Investigating both types of input states thus offers a thorough assessment for the efficiency of the PS-TMSFS resource.

Our results provide insight into the role of PS-TMSFS in improving performance of the CV quantum teleportation. The paper is organized as follows: in Section~\ref{WCF}, we describe the scheme for preparing PS-TMSFS. The Wigner characteristic function and the success probability associated with photon-subtraction are evaluated for the resource state. In Section~\ref{fidelity}, teleportation fidelities for input Gaussian and non-Gaussian states are obtained . Finally in section~\ref{conclusion}, we draw our conclusion based on the results obtained. In Appendix~\ref{App_A}, we provide a brief overview of continuous-variable systems and their phase-space representations. Appendix~\ref{App_B} presents matrices and coefficients appearing in the descriptions of the Wigner characteristic function, success probability and teleportation fidelity.

\section{WIGNER CHARACTERISTIC FUNCTION OF PS-TMSFS}
\label{WCF}
\begin{figure}[H]
	\includegraphics[trim=1pt 3pt 1pt 3pt,clip,width=0.5\textwidth]{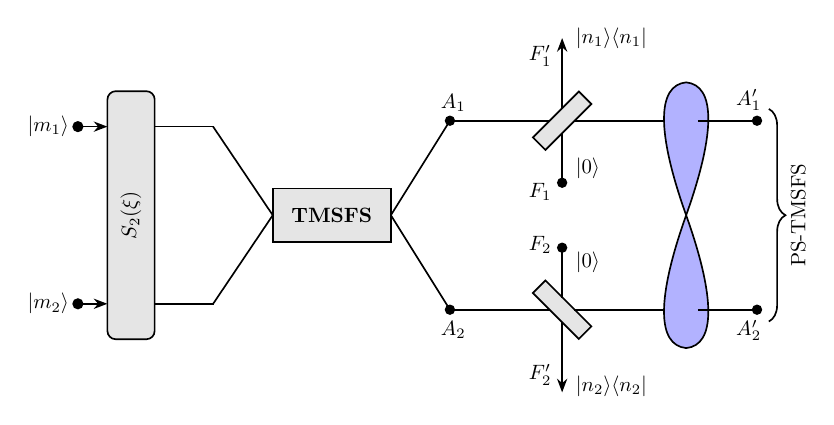}
	\caption{Schematic representation for preparing PS-TMSFS. The input modes of Fock states $\ket{m_1}$ and $\ket{m_2}$ undergo a two-mode squeezing operation $S_2(\xi)$ to produce TMSFS. The modes $A_1$ and $A_2$ of TMSFS are mixed with ancillary vacuum modes $F_1$ and $F_2$ via beam-splitters with transmissivities $T_1$ and $T_2$, respectively. Photon-subtraction is implemented via conditional measurements using photon-number-resolving-detectors (PNRDs) on the ancillary modes $F_1$ and $F_2$, described by the positive-operator-valued-measures (POVMs) $\ket{n}\bra{n}$. Successful detection of $\ket{n_1}\bra{n_1}$ and $\ket{n_2}\bra{n_2}$ in the ancillary modes corresponds to the subtraction of $n_1$ and $n_2$ photons from modes $A_1$ and $A_2$, respectively, thereby generating PS-TMSFS.}
	\label{Fig1}
\end{figure}

We first illustrate the experimental set-up for generating the PS-TMSFS in Fig.~\ref{Fig1}.
We begin with an uncorrelated two-mode Fock state which serves as the input for the two-mode squeezing operation. This state provides the basis for constructing the two-mode squeezed Fock state (TMSFS) $S_2(\xi)\ket{m_1, m_2}$ whose non-classical correlations are subsequently modified via photon subtraction. The initial two-mode Fock state can be written as
\begin{equation}
	|\psi\rangle_{m_1 m_2} = |m_1,m_2\rangle
\end{equation}
with $m_1$ and $m_2$ photons in the first and second modes, respectively. Since the two-mode Fock state with higher photon numbers ($n\ge 1$) is a non-Gaussian state, its Wigner characteristic function can be expressed in terms of Laguerre polynomials using Eq.~\eqref{Lag} as
\begin{align}
	\chi_{|m_1, m_2\rangle}(\tau_1,\sigma_1,\tau_2,\sigma_2)
	&	= L_{m_1}\!\left( \frac{\tau_1^2 + \sigma_1^2}{2} \right)
	L_{m_2}\!\left( \frac{\tau_2^2 + \sigma_2^2}{2} \right) \nonumber \\ &	\exp \!\left( -\frac{\tau_1^2 + \sigma_1^2 + \tau_2^2 + \sigma_2^2}{4} \right).
	\label{eqwigner2}
\end{align}
To obtain the TMSFS, the two-mode squeezing operator $S_2(\xi)$ associated with a non-degenerate parametric down-conversion process is applied to the initial two-mode Fock state as
\begin{equation}
	|\psi\rangle_{A_1 A_2} = S_2(\xi)|\psi\rangle_{m_1 m_2}, 
\end{equation}
where $S_2(\xi)$ is the two-mode squeezing operator defined in Eq.~\eqref{TMSO} of Appendix~\ref{App_A}. Using the disentangled form of the two-mode squeezing operator \eqref{TMSV_disentangle}, an explicit Fock-basis representation of the TMSFS can be written as
	\begin{align}
		S_2(\xi)\ket{m_1,m_2}
		& =	\sum_{k=0}^{\infty}\,\,
		\sum_{l=0}^{\min(m_1, m_2)}	
		\frac{(-1)^l}{l!\,k!}
		(\tanh\xi)^{k+l}
		(\sech\xi)^{m_1+m_2-2l+1}
		\nonumber\\
		&\times
		\frac{
			\sqrt{
				m_1!m_2!
				(m_1-l+k)!
				(m_2-l+k)!
			}
		}{
			(m_1-l)!(m_2-l)!
		}
		\nonumber\\
		&\times
		\ket{m_1-l+k,m_2-l+k}.
		\label{eq:TMSFS_Fock}
	\end{align}
	The two-mode squeezing operator creates or annihilates photons only in correlated pairs, one in each mode. Consequently,
	\[
	[\hat n_1-\hat n_2,S_2(\xi)]=0,
	\]
	that means the photon-number difference is conserved.
	In the special case of $m_1=m_2=0$, Eq.~\eqref{eq:TMSFS_Fock} reduces to the well-known two-mode squeezed vacuum state,
	\begin{equation}
		S_2(\xi)\ket{0,0}
		=
		\sech\xi
		\sum_{k=0}^{\infty}
		(\tanh\xi)^k
		\ket{k,k}.
	\end{equation}
	Furthermore, in the weak-squeezing limit $(\xi\ll1)$, Eq.~\eqref{eq:TMSFS_Fock} reduces to
	\begin{align}
		S_2(\xi)\ket{m_1,m_2}
		&	\simeq
		\ket{m_1,m_2} \nonumber \\&
		+
		\xi
		\sqrt{(m_1+1)(m_2+1)}
		\ket{m_1+1,m_2+1} \nonumber\\ &
		-
		\xi
		\sqrt{m_1m_2}
		\ket{m_1-1,m_2-1}
		+
		O(\xi^2).
		\label{eq:Weak_TMSFS}
	\end{align}
	Eq.~\eqref{eq:Weak_TMSFS} shows that for weak squeezing, the initial Fock state remains the dominant contributor, while the two-mode squeezing operation introduces small correlated photon-pair creation and annihilation amplitudes proportional to $\sqrt{(m_1+1)(m_2+1)}$ and $\sqrt{m_1m_2}$, respectively.

	To further elucidate the physical mechanism underlying the considered scheme, we consider the symmetric case $n_1=n_2=1$ and $m_1=m_2=1$ in the limiting case $T\rightarrow1$. In this limit, the effects of the beam splitters vanish, and the resulting non-Gaussian state takes the form of a two-mode squeezed superposition of Fock state
\begin{align*}
	a_1a_2S_2(\xi)|1,1\rangle &=S_2(\xi)
	\Big[
	a_1a_2\cosh^2\xi
	+(a_1a_1^\dagger+a_2^\dagger a_2)
	\sinh\xi\cosh\xi  \\ &
	+a_1^\dagger a_2^\dagger\sinh^2\xi
	\Big]|1,1\rangle \\ 
&=
	\frac{S_2(\xi)}{1-\lambda^2}
	\left[
	|0,0\rangle
	+3\lambda |1,1\rangle
	+2\lambda^2 |2,2\rangle
	\right].
\end{align*}
with $
\lambda=\tanh\xi $.
This shows that photon subtraction changes the relative weights of the correlated Fock-state components, producing a non-Gaussian superposition that contributes to the two-mode squeezing of the overall state. This contribution adds to the squeezing generated by $S_2(\xi)$ and can therefore enhance the squeezing properties of the considered scheme. A similar Fock-basis interpretation for squeezing distillation in the single-mode case was discussed in Ref.~\cite{Fiur2025}.

Under the two-mode squeezing transformation, the Wigner characteristic function given in Eq.~\eqref{eqwigner2} evolves as $
\chi(\Lambda)\rightarrow\chi(S_2(\xi)\Lambda)$ 
\begin{align}
	\label{WCF_TMSFS}
	&\chi_{A_1A_2}(\Lambda)
	= \exp\!\Bigg[
	-\frac{1}{4}\Big(
	\cosh(2\xi)(\tau_1^2+\sigma_1^2+\tau_2^2+\sigma_2^2)
	\nonumber\\
	&\qquad\qquad
	-2\sinh(2\xi)(\tau_1\tau_2-\sigma_1\sigma_2)
	\Big)\Bigg]
	\nonumber\\[4pt]
	&\times
	\frac{1}{2^{m_1}m_1!}
	\frac{\partial^{m_1}}{\partial u_1^{m_1}}
	\frac{\partial^{m_1}}{\partial v_1^{m_1}}
	\Bigg[
	2u_1v_1
	+u_1\Big(
	(\cosh \xi\,\tau_1-\sinh \xi\,\tau_2)
	\nonumber\\
	&\qquad
	+i(\cosh \xi\,\sigma_1+\sinh \xi\,\sigma_2)
	\Big)
	-v_1\Big(
	(\cosh \xi\,\tau_1-\sinh \xi\,\tau_2)
	\nonumber\\
	&\qquad
	-i(\cosh \xi\,\sigma_1+\sinh \xi\,\sigma_2)
	\Big)
	\Bigg]_{
		u_1=v_1=0 
	}
	\nonumber\\[6pt]
	&\times
	\frac{1}{2^{m_2}m_2!}
	\frac{\partial^{m_2}}{\partial u_2^{m_2}}
	\frac{\partial^{m_2}}{\partial v_2^{m_2}}
	\Bigg[
	2u_2v_2
	+u_2\Big(
	(-\sinh \xi\,\tau_1+\cosh \xi\,\tau_2)
	\nonumber\\
	&\qquad
	+i(\sinh \xi\,\sigma_1+\cosh \xi\,\sigma_2)
	\Big)
	-v_2\Big(
	(-\sinh \xi\,\tau_1+\cosh \xi\,\tau_2)
	\nonumber\\
	&\qquad
	-i(\sinh \xi\,\sigma_1+\cosh \xi\,\sigma_2)
	\Big)
	\Bigg]_{
		u_2=v_2=0 
	}.
\end{align}
Next, the modes $A_1$ and $A_2$ of the TMSFS are combined with ancilla vacuum modes $F_1$ and $F_2$ using beam-splitters of transmissivities $T_1$ and $T_2$, respectively. We represent the modes $A_1$ and $A_2$ by the quadrature	operators $(\hat{q}_1,\hat{p}_1)$ and $(\hat{q}_2,\hat{p}_2)$, and the ancilla modes $F_1$ and $F_2$ by the quadrature operators $(\hat{q}_3,\hat{p}_3)$ and $(\hat{q}_4,\hat{p}_4)$, respectively. Prior to the beam-splitter operation, the Wigner characteristic function of the four-mode system can be written as
\begin{equation}
	\chi_{F_1A_1A_2F_2}(\Lambda) =
	\chi_{A_1A_2}(\Lambda)~
	\chi_{\ket{0}}(\Lambda_3)~\chi_{\ket{0}}(\Lambda_4),
\end{equation}
where $\chi_{\ket{0}}(\Lambda_i), (i=3, 4)$ is the Wigner characteristic function of the vacuum state. After the beam-splitter operation, the four modes get entangled by mixing the modes by the two beam-splitters, collectively represented by the symplectic transformation matrix $B(T_1, T_2)=B_{A_1F_1}(T_1)\oplus B_{A_2F_2}(T_2)$ and is evolved as
\begin{equation}
	\chi_{F'_1A'_1A'_2F'_2}(\Lambda) =
	\chi_{F_1A_1A_2F_2}(B(T_1, T_2)^{-1}\Lambda). 
\end{equation}
The modes $A_1'$ and $A_2'$ represent the output modes obtained after the beam-splitter interaction with the ancillary modes $F_1$ and $F_2$, respectively. The modes $F'_1$ and $F'_2$ are measured with PNRDs represented by the projection operators
$|n_1\rangle\langle n_1|$ and $|n_2\rangle\langle n_2|$, respectively. When an equal number of photons is subtracted from each mode, i.e., $n_1 = n_2 = n$, the process corresponds to symmetric photon-subtraction, yielding the symmetric PS-TMSFS and when the photon numbers differ, i.e., $n_1 \neq n_2$, the process corresponds to asymmetric photon-subtraction, leading to the asymmetric PS-TMSFS.

The unnormalized Wigner characteristic function can be written as
\begin{align} 
	\widetilde{\chi}_{A'_1A'_2}^{\mathrm{PS}} (\Lambda)
	=&
	\frac{1}{(2\pi)^2}
	\int d^2 \Lambda_3 \, d^2 \Lambda_4 \;
	\underbrace{\chi_{F_1' A_1' A_2' F_2'}(\Lambda)}_{\text{joint four-mode characteristic function}}\nonumber \\ & \qquad
	\underbrace{\chi_{|n_1\rangle}(\Lambda_3)}_{\text {Fock-state projection}} \quad
	\underbrace{\chi_{|n_2\rangle}(\Lambda_4)}_{\text{ Fock-state projection}}, \label{Int}
\end{align}
where $\chi_{|n\rangle}(\Lambda)$ is the Wigner characteristic function of Fock state $|n\rangle$ which can be written in terms of Laguerre polynomial as given in Eq.~\eqref{TMSO} of the Appendix~\ref{App_A}. Integrating Eq.~\eqref{Int} yields
\begin{align}
	\label{eq11}
	\widetilde{\chi}_{A'_1A'_2}^{\mathrm{PS}} (\Lambda) = \frac{\hat{F}}{a_0}\exp(\Lambda^T M_1 \Lambda + u^T M_2  \Lambda + u^T M_3 u),
\end{align} 
where $a_0=\beta^2-\alpha^2t_1^2t_2^2$ and the parameters $\alpha$, $\beta$, $t_1$ and $t_2$ are given in Appendix \ref{App_B}. The column vectors $\Lambda$ and $u$ are defined as
$(\tau_1, \sigma_1, \tau_2, \sigma_2)^T$ and $(u_1, v_1, u_2, v_2, u'_1, v'_1, u'_2, v'_2)^T$, 	respectively, and the matrices $M_1$, $M_2$ and $M_3$ are given by Eqs.~\eqref{M_1}, \eqref{M_2} and \eqref{M_3} of the Appendix \ref{App_B}. Further, the differential operator $\hat{F}$ is defined as
\begin{align*}
	\hat{F}  = &\;
	\frac{2^{-(m_1+m_2+n_1+n_2)}}{m_1!m_2!n_1!n_2!}
	\;
	\frac{\partial^{m_1}}{\partial u_1^{m_1}}
	\;
	\frac{\partial^{m_1}}{\partial v_1^{m_1}}
	\;
	\frac{\partial^{m_2}}{\partial u_2^{m_2}}
	\;
	\frac{\partial^{m_2}}{\partial v_2^{m_2}} \\[6pt]
	&\quad
	\frac{\partial^{n_1}}{\partial {u'_1}^{n_1}}
	\;
	\frac{\partial^{n_1}}{\partial {v'_1}^{n_1}}
	\;
	\frac{\partial^{n_2}}{\partial {u'_2}^{n_2}}
	\;
	\frac{\partial^{n_2}}{\partial {v'_2}^{n_2}}
	\Bigg|_{\substack{
			u_1=v_1=u_2=v_2=0 \\
			u'_1=v'_1=u'_2=v'_2=0
	}}
\end{align*}

	The normalized Wigner characteristic function corresponding to Eq.~\eqref{eq11} is given by
	\begin{eqnarray}
		\chi_{A'_1A'_2}^{\mathrm{PS}}(\Lambda)
		=
		\left(P^{\mathrm{PS}}\right)^{-1}
		\widetilde{\chi}_{A'_1A'_2}^{\mathrm{PS}}(\Lambda)
		\label{Normalise}
	\end{eqnarray}
	where $P^{\mathrm{PS}}$ is the normalization factor determined by the condition $\chi_{A'_1A'_2}^{\mathrm{PS}}(0)=1$. It is therefore given by
	\begin{equation}
		P^{\mathrm{PS}}
		=
		\left.
		\widetilde{\chi}_{A'_1A'_2}^{\mathrm{PS}}(\Lambda)
		\right|_{\tau_1=\sigma_1=\tau_2=\sigma_2=0}
		=
		\frac{\hat{F}}{a_0}
		\exp\left(u^{T}M_3u\right).
	\end{equation}
	Here $P^{\mathrm{PS}}$ physically represents the success probability of the photon-subtraction operation. To evaluate the feasibility of this process, the behavior of $P^{\mathrm{PS}}$ is analyzed for different photon-subtraction configurations as follows.
\begin{figure}[H]
	\centering
	\includegraphics[trim=0pt 0pt 0pt 0pt,clip,width=0.5\textwidth]{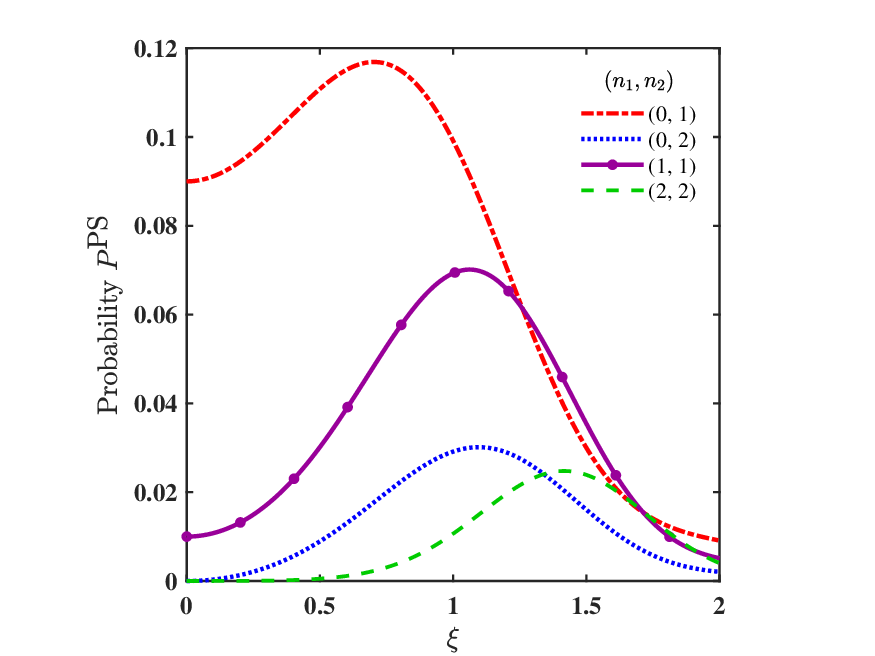}
	\caption{Success probability $P^{\mathrm{PS}}$ as a function of the squeezing parameter $\xi$ for different photon-subtraction configurations $(n_1,n_2)=(0,1)$, $(0,2)$, $(1,1)$ and $(2,2)$ of the PS-TMSFS channel at fixed transmissivity $T_1 = T_2 = 0.9$ and with the initial Fock state $(m_1,m_2) = (1,1)$. This plot illustrates the dependence of the success probability on the squeezing parameter and the photon-subtraction numbers.} \label{Plot2}
\end{figure}
\begin{figure}[H]
	\centering
	\includegraphics[trim=0pt 0pt 0pt 0pt,clip,width=0.5\textwidth]{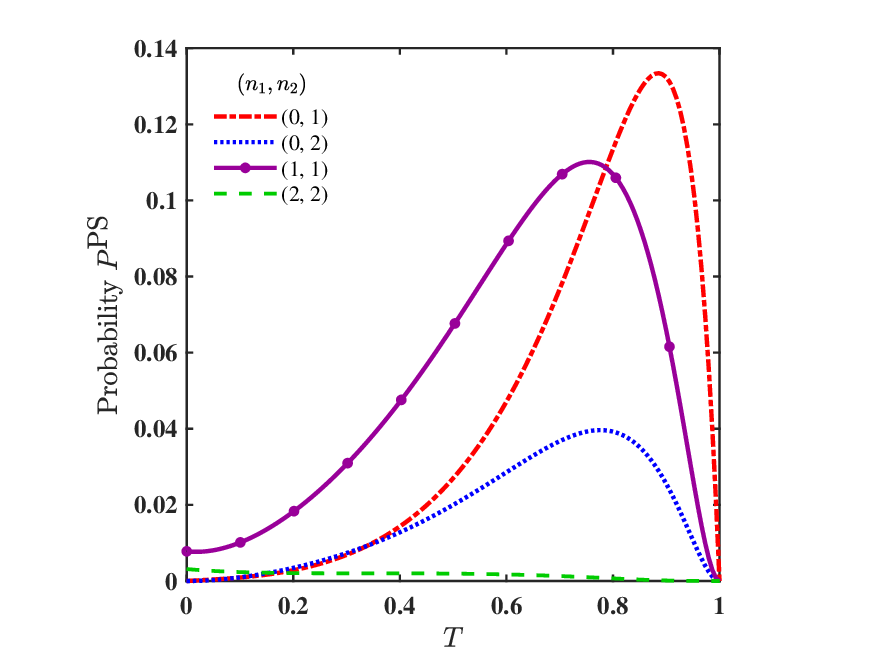}
	\caption{Success probability $P^{\mathrm{PS}}$ as a function of the transmissivity $T$ ($T_1 = T_2 = T$) for different photon-subtraction configurations $(n_1,n_2)$ at fixed squeezing parameter $\xi = 0.8$ and with the initial Fock state $(m_1,m_2) = (1,1)$. This plot illustrates the dependence of the success probability on the transmissivity and the numbers of photons subtracted.} \label{Plot3}
\end{figure}
The success probability exhibits a strong dependence on both the squeezing parameter and the beam-splitter transmissivity, as shown in Figs.~\ref{Plot2} and~\ref{Plot3}. In Fig.~\ref{Plot2}, where the transmissivity is fixed at $T = 0.9$, the success probability initially increases with the squeezing parameter $\xi$, reaches a maximum, and then gradually decreases for higher values of $\xi$.  Among the different configurations, the asymmetric single-photon subtraction case $(0,1)$ yields the highest success probability, while higher-order subtractions exhibit significantly lower probabilities due to the reduced likelihood of multi-photon detection events.

In Fig.~\ref{Plot3}, where the squeezing parameter is fixed, the success probability varies non-monotonically with the transmissivity $T$. For all photon-subtraction configurations, the probability increases with $T$, rises to a peak beyond an intermediate value, and then decreases rapidly as $T \rightarrow 1$. This is consistent with the fact that very high transmissivity suppresses photon-subtraction events, thereby reducing the success probability. Similar to the previous case, lower-order photon-subtraction leads to higher success probability compared to higher-order processes.

Overall, the results indicate that both the squeezing parameter and beam-splitter transmissivity must be carefully optimized to maintain a balance between non-Gaussian state preparation and detection efficiency. The interplay between these parameters is crucial in determining the feasibility of photon-subtracted resource states for CV quantum teleportation.

Beyond the success probability analysis, the normalized Wigner characteristic function in Eq.~\eqref{Normalise} provides a unified framework to obtain different relevant quantum states as limiting cases. By considering $T_1 \rightarrow 1$ and $T_2 \rightarrow 1$ in the symmetric PS-TMSFS with $n_1=n_2=0$, we obtain the Wigner characteristic function of the ideal TMSFS. Similarly $m_1=m_2=0$ renders the Wigner characteristic function of PS-TMSV state and $\xi=0$ results in photon-subtracted two-mode Fock state. 
\section{Teleportation Fidelity}
\label{fidelity}
To provide a quantitative assessment for the entangled resource for quantum information processing, we analyze its performance in CV quantum teleportation. Throughout this work, the Braunstein-Kimble (BK) protocol \cite{Braunstein1998} is considered as the teleportation scheme. 

The maximum fidelity achievable for teleporting a coherent state without using a shared entangled channel is \(1/2\)~\cite{Braunstein2000TF,Braunstein2001}. Consequently, any fidelity exceeding \(1/2\) indicates successful CV quantum teleportation, whereas unit value of fidelity corresponds to an ideal state transfer. In addition, a fidelity exceeding the no-cloning threshold of $2/3$ is generally required for secure quantum teleportation \cite{Cerf2000,Grosshans2001}, ensuring that no unauthorized party can avail a copy of the teleported state. In this work, we investigate the performance of the PS-TMSFS as an entangled resource. 
As all the relevant states are conveniently represented in phase space, the analysis can be carried out entirely within the characteristic function formalism.

In the BK protocol, the sender (Alice) and the receiver (Bob) shared a bipartite entangled resource, 
denoted by the density operator $\rho_{A'_1A'_2}$ with the corresponding Wigner characteristic function $\chi_{A'_1A'_2}^{\text{PS}}(\Lambda)$. Alice is provided with an unknown input state $\rho_{\mathrm{in}}$ characterized by the Wigner function $\chi_{\mathrm{in}}(\Lambda_{\mathrm{in}})$, which is to be teleported to Bob. In the protocol, the input mode and Alice's component of the entangled resource are mixed at a balanced beam-splitter. Subsequently, homodyne measurements of the conjugate quadratures are performed and the measurement outcomes are transmitted to Bob through a classical communication channel, enabling him to apply an appropriate displacement operator on his mode. As a result, Bob obtains the output state $\rho_{\mathrm{out}}$, which represents the teleported version of the original input state.

In the framework of phase-space method, the Wigner characteristic function of the output state can be expressed as \cite{Marian2006}
\begin{equation}
	\chi_{\mathrm{out}}(\tau_2,\sigma_2)=
	\chi_{\mathrm{in}}(\tau_2,\sigma_2)~
	\chi_{\mathrm{res}}(\tau_2,-\sigma_2,\tau_2,\sigma_2).
\end{equation}
The quality of the teleportation process is quantified through the teleportation fidelity, defined as the overlap between the input state and the teleported output state as
\begin{equation}
	F=\mathrm{Tr}\left[\rho_{\mathrm{in}}~\rho_{\mathrm{out}}\right].
\end{equation}
Reformed in terms of the Wigner characteristic functions, the fidelity reads as \cite{Chizhov2002}
\begin{equation}
	F=	\frac{1}{2\pi}
	\int d^2\Lambda_{\mathrm{in}}~
	\chi_{\mathrm{in}}(\Lambda_{\mathrm{in}})~
	\chi_{\mathrm{out}}(-\Lambda_{\mathrm{in}}).
	\label{F_B}
\end{equation}
In the following subsections, the fidelity expressions \eqref{F_B} are computed using the Wigner characteristic functions of the input states and the PS-TMSFS resource. The resulting fidelity is employed as primary figure of merit to evaluate the ability of the resource for teleporting different types of Gaussian and non-Gaussian input states. 

\subsection{Teleportation of Gaussian Input States}
	We investigate the teleportation of Gaussian input states using the PS-TMSFS as an entangled resource. Coherent and squeezed states, being fundamental Gaussian states, provide a suitable benchmark for evaluating the performance of the teleportation protocol. In this subsection, we consider the teleportation of coherent and squeezed vacuum states and evaluate the corresponding fidelities.
\subsubsection{Coherent State}
We first evaluate the fidelity for the teleportation of a single-mode coherent state using the PS-TMSFS as the entangled resource. Using the Wigner characteristic functions of the coherent state \eqref{WCF_CS} and PS-TMSFS \eqref{Normalise} in Eq.~\eqref{F_B}, the teleportation fidelity is derived as
\begin{align}
	{F	= \frac{\hat{F}}{d_0}   \;	\exp \left(u^TM_4u \right)}
	\label{FCS}
\end{align}
where $\hat{F}$ denotes the differential operator and $d_0=2\beta (\beta - \alpha t_1t_2)$. The matrix $M_4$ is given in Eq.~\eqref{M_4} of Appendix \ref{App_B}. The above fidelity expression specifies the effects of the squeezing parameter, beam-splitter transmissivities, and photon-subtraction operations used in the resource state. Consequently, it provides an insightful framework for investigating how these parameters influence the fidelity. We first analyze the behavior of the teleportation fidelity as a function of the squeezing parameter  for different photon-subtraction configurations of the PS-TMSFS resource. We have also included the TMSV state as a benchmark for comparison. Comparing the proposed resource with this standard reference highlights its relative performance and clearly illustrates the improvement over the conventional TMSV state.           

\begin{figure}[H]
	\centering
	\includegraphics[trim=0pt 0pt 0pt 0pt,clip,width=0.5\textwidth]{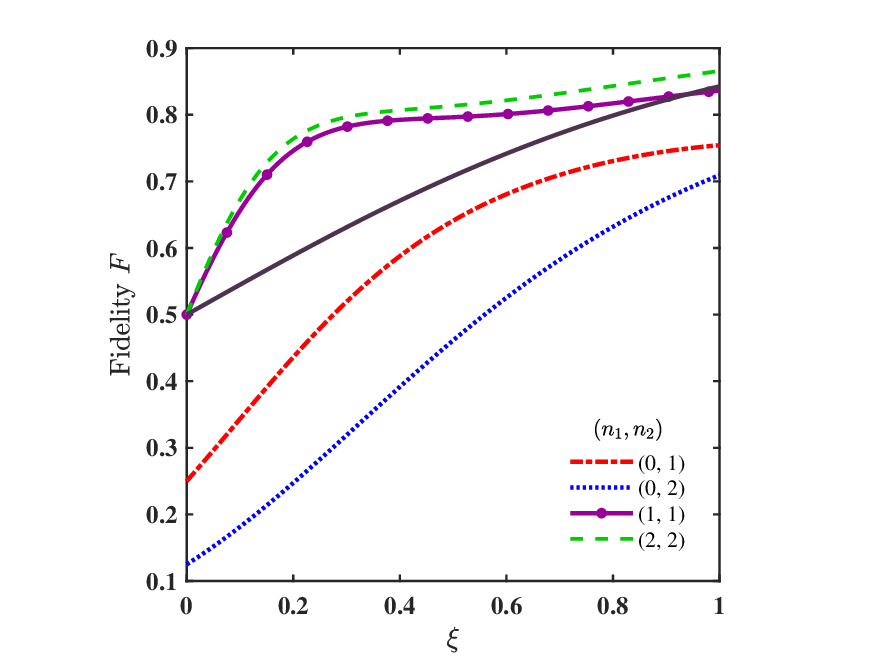}
	\caption{Teleportation fidelity $F$ for the input coherent state as a function of the squeezing parameter $\xi$ for different photon-subtraction configurations $(n_1,n_2)=(0,1)$, $(0,2)$, $(1,1)$ and $(2,2)$. The underlying Fock state in PS-TMSFS is fixed at $(m_1,m_2)=(1,1)$ with the beam-splitter transmissivities $T_1=T_2=0.9$. The solid black line corresponds to the teleportation fidelity of input coherent state using the two-mode squeezed vacuum (TMSV) resource.
	}
	\label{R_coh}
\end{figure}

Fig.~\ref{R_coh} shows the teleportation fidelity as a function of the squeezing parameter $\xi$ for different photon-subtraction configurations of the PS-TMSFS resource. For all cases considered, the fidelity exhibits a pronounced dependence on the squeezing parameter, indicating that the teleportation performance is strongly influenced by the degree of squeezing present in the resource state.

Across the different configurations shown in Fig.~\ref{R_coh}, the symmetric photon-subtraction cases $(1,1)$ and $(2,2)$ provide a significant enhancement in the teleportation fidelity over the Gaussian TMSV resource in the low and intermediate squeezing regimes. Starting from the classical limit of $1/2$ at zero squeezing, the fidelities for both the configurations increase rapidly with the squeezing parameter and saturate for larger values of $\xi$. Throughout the entire parameter range, the $(2,2)$ configuration consistently yields the highest fidelity while the $(1,1)$ configuration closely follows with only a marginally lower performance. In contrast, the asymmetric photon-subtraction configurations exhibit considerably lower fidelities. Although the $(0,1)$ configuration increases monotonically with the squeezing parameter, its fidelity remains well below than the TMSV curve. The $(0,2)$ configuration exhibits the lowest efficiency among all the resources considered. These results demonstrate that the teleportation fidelity depends strongly on both the squeezing strength and the photon-subtraction configuration with the symmetric two-mode photon-subtraction providing the most pronounced enhancement in the teleportation performance.

After examining the influence of the squeezing parameter, we investigate the role of the beam-splitter transmissivity used in the photon-subtraction process. Since the transmissivity determines the strength of the conditional subtraction operation, it provides an additional degree for optimizing the teleportation fidelity of the resource state.
\begin{figure}[H]
	\centering
	\includegraphics[trim=0pt 0pt 0pt 0pt,clip,width=0.5\textwidth]{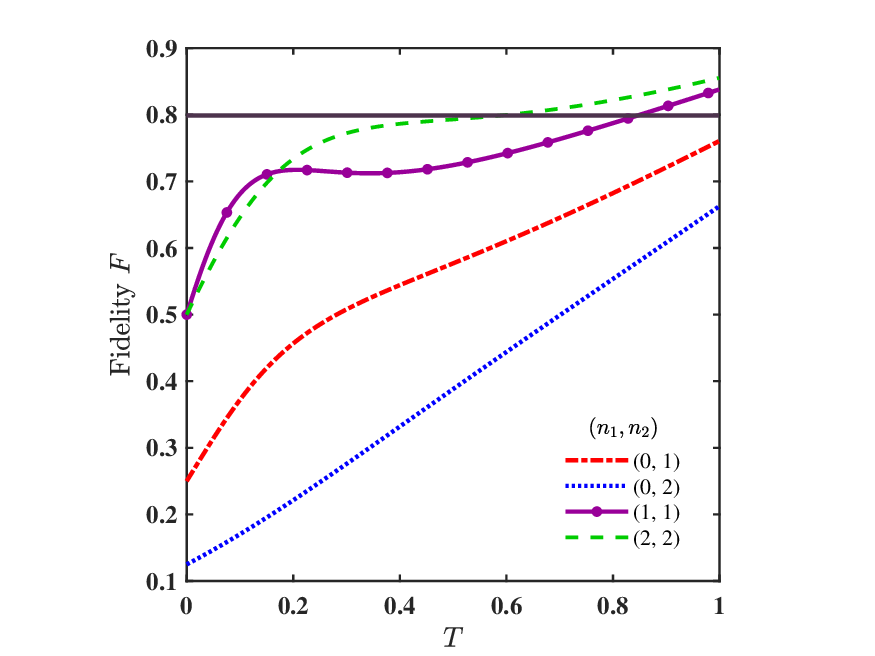}
	\caption{Teleportation fidelity $F$ as a function of the beam-splitter transmissivity $T$ ($T_1=T_2=T$) for different photon-subtraction configurations $(n_1,n_2)$ of the PS-TMSFS channel. The underlying Fock state is chosen as $(m_1,m_2)=(1,1)$	with the squeezing parameter $\xi=0.8$. The solid black line represents the fidelity of the Gaussian TMSV resource.} 
	\label{T_coh}
\end{figure}
	Fig.~\ref{T_coh} shows the teleportation fidelity as a function of the beam-splitter transmissivity $T$ employed in the photon-subtraction process. The Gaussian TMSV resource, represented by the solid black line, is not undergoing any photon-subtraction process. It is therefore independent of the beam-splitter transmissivity and acts as a fixed reference throughout the figure. Among all the photon-subtraction configurations, the symmetric $(2,2)$ case yields the highest fidelity throughout most of the transmissivity range. Starting from the classical limit of $1/2$ at $T=0$, its fidelity increases rapidly for small transmissivities and then exhibits a gradual increase for larger values of $T$, attaining the maximum fidelity value of 0.86. The $(1,1)$ configuration shows a similar dependence on the beam-splitter transmissivity. It exhibits a pronounced dependence on $T$ with the fidelity increasing rapidly at low transmissivities before approaching saturation as $T\rightarrow 1$. Although the Gaussian TMSV resource outperforms the $(1,1)$ configuration over a substantial portion of the considered transmissivity range, the $(1,1)$ curve crosses the TMSV plot at approximately $T\simeq0.85$ and yields a higher fidelity for larger transmissivities. In contrast, the asymmetric photon-subtraction configurations exhibit consistently lower fidelities. Both $(0,1)$ and $(0,2)$ increase monotonically with transmissivity; however, their fidelities remain below that of the Gaussian TMSV resource over the entire parameter range. Between the two asymmetric cases, the $(0,1)$ configuration consistently outperforms the $(0,2)$ configuration, while the latter exhibits the lowest fidelity among all the considered configrations. In general, these results demonstrate that the symmetric photon-subtraction is substancially more effective than the asymmetric subtraction for enhancing the teleportation fidelity with the (2,2) configuration providing the best overall performance.
\subsubsection{Squeezed Vacuum State}

Next we calculate the fidelity for teleporting a single-mode squeezed vacuum state. Unlike a coherent state, a squeezed vacuum state possesses reduced fluctuation in one quadrature at the cost of enhanced fluctuation in the conjugate quadrature, rendering it a more sensitive probe of quantum correlations present in the teleportation channel. Consequently, the corresponding teleportation fidelity provides a deeper insight into the potential of the PS-TMSFS resource to preserve the non-classical features of the input squeezed state $\ket{\epsilon}=S(\epsilon)\ket{0}$.

Using the Wigner characteristic function of the squeezed vacuum state given in Eq.~\eqref{WCF_SS} together with that of the PS-TMSFS, the fidelity is evaluated as 	
\begin{align}
	F=	\frac{\hat{F}}{e'_0}
	\exp\left(u^{T}M_5u\right),
	\label{FSS}
\end{align}
where $e'_0= \sqrt{e_0}\,(\beta-\alpha t_1 t_2)$, $\hat{F}$ is the differential operator and the matrix $M_5$ is given in Eq.~\eqref{M_5}. The fidelity can then be studied as a function of the squeezing ($\xi$), the beam-splitter tranmittivity ($T$) as well as the input squeezing parameter ($\epsilon$).
\begin{figure}[H]
	\centering
	\includegraphics[trim=0pt 0pt 0pt 0pt,clip,width=0.5\textwidth]{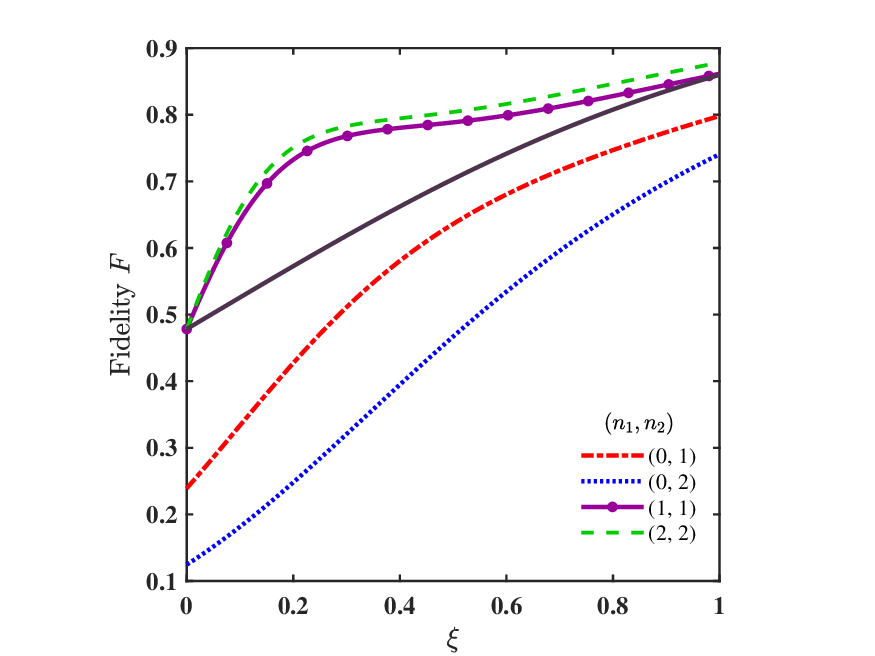}
	\caption{Dependence of the teleportation fidelity $F$ on the squeezing parameter $\xi$ for photon-subtraction configurations $(n_1,n_2)=(0,1)$, $(0,2)$, $(1,1)$ and $(2,2)$ with the source Fock state $(m_1,m_2)=(1,1)$ and beam-splitter transmissivities $T_1=T_2=0.9$. The squeezing of the input squeezed vacuum state is set to $\epsilon=0.3$. The solid black line represents the teleportation fidelity for an input squeezed vacuum state using the TMSV channel.	
	}
	\label{R_seq}
\end{figure}
Fig.~\ref{R_seq} depicts the variation of the teleportation fidelity with the resource squeezing parameter $\xi$ for an input squeezed vacuum state.
	Similar to the coherent state, the teleportation fidelity depends strongly on the squeezing parameter with the symmetric photon-subtraction configurations providing the higher fidelities over the entire range considered. Among all the investigated resources, the symmetric $(2,2)$ configuration consistently yields the highest teleportation fidelity while the $(1,1)$ configuration closely follows it with only a marginally lower performance. Both the configurations exhibit a rapid increase in fidelity in the low-squeezing regime and increase steadily as the squeezing parameter increases. The Gaussian TMSV resource also exhibits a monotonic increase with $\xi$. While its fidelity approaches that of the $(1,1)$ photon-subtraction case at higher squeezing, it remains below that of the $(2,2)$ configuration across the entire considered range. In contrast, the asymmetric photon-subtraction configurations yield significantly lower fidelities. 
	Although the $(0,1)$ configuration increases monotonically with $\xi$, it remains below that of the Gaussian TMSV resource and the $(0,2)$ configuration consistently exhibits the lowest fidelity over the entire range. 
	
	The above results confirm that the symmetric photon-subtraction provides the most effective improvement in the teleportation fidelity for squeezed vacuum state. Next, we investigate the effect of the beam-splitter transmissivity.

\begin{figure}[H]
	\centering
	\includegraphics[trim=0pt 0pt 0pt 0pt,clip,width=0.5\textwidth]{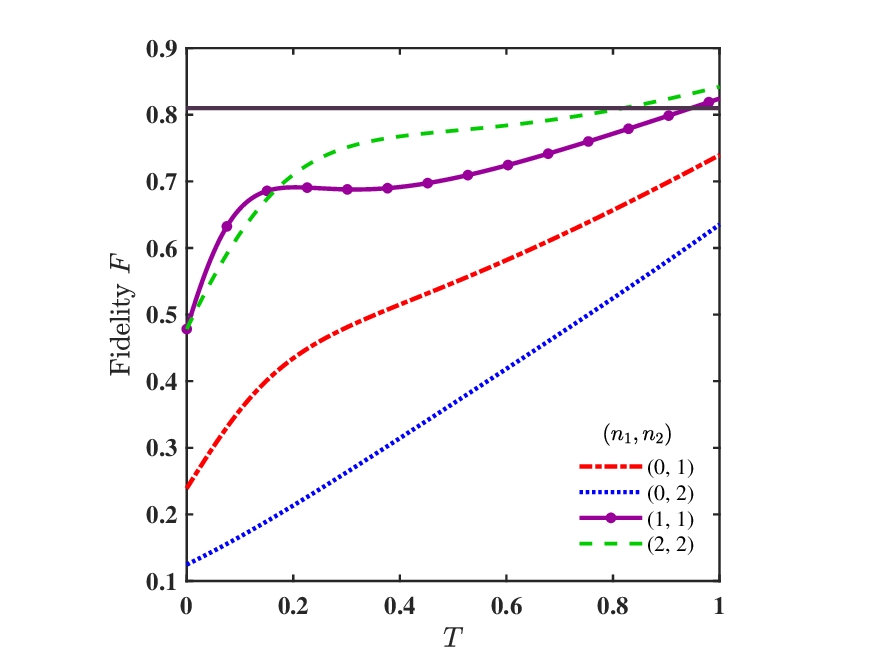}
	\caption{Dependence of the teleportation fidelity $F$ on beam-splitter transmittance $T$ for input squeezed vacuum state with different photon-subtraction operations given by $(n_1,n_2)$. The resource state is prepared with $(m_1,m_2)=(1,1)$ and squeezing parameter $\xi=0.8$, while the input squeezed state is fixed at $\epsilon=0.3$. The solid black line represents the Gaussian TMSV resource.
	}
	\label{T_seq}
\end{figure}
Fig.~\ref{T_seq} illustrates the dependence of the teleportation fidelity for the input squeezed vacuum state on the beam-splitter transmissivity employed in the preparation of the PS-TMSFS resource. The transmissivity dependence also closely resembles that obtained for the coherent state input. The solid black line representing the Gaussian TMSV resource remains constant throughtout the entire range of $T$ and provides the highest fidelity up to $T \simeq 0.8$, beyond which it is surpassed by the symmetric $(2,2)$ configuration. The $(1,1)$ configuration also increases rapidly at low transmissivities, followed by a broad region of near-constant behavior over the intermediate range, and approaches that of the TMSV resource only as $T\rightarrow1$. Unlike this, the asymmetric $(0,1)$ configuration increases monotonically with $T$ but remains below the Gaussian TMSV resource throughout the entire range, whereas the $(0,2)$ configuration yields the lowest fidelity. These observations further confirm the advantage of the symmetric photon-subtraction for teleporting Gaussian input states. After examining the influence of the resource parameters, we investigate the effect of the input squeezing parameter $\epsilon$ on the teleportation fidelity.

\begin{figure}[H]
	\centering
	\includegraphics[trim=0pt 0pt 0pt 0pt,clip,width=0.5\textwidth]{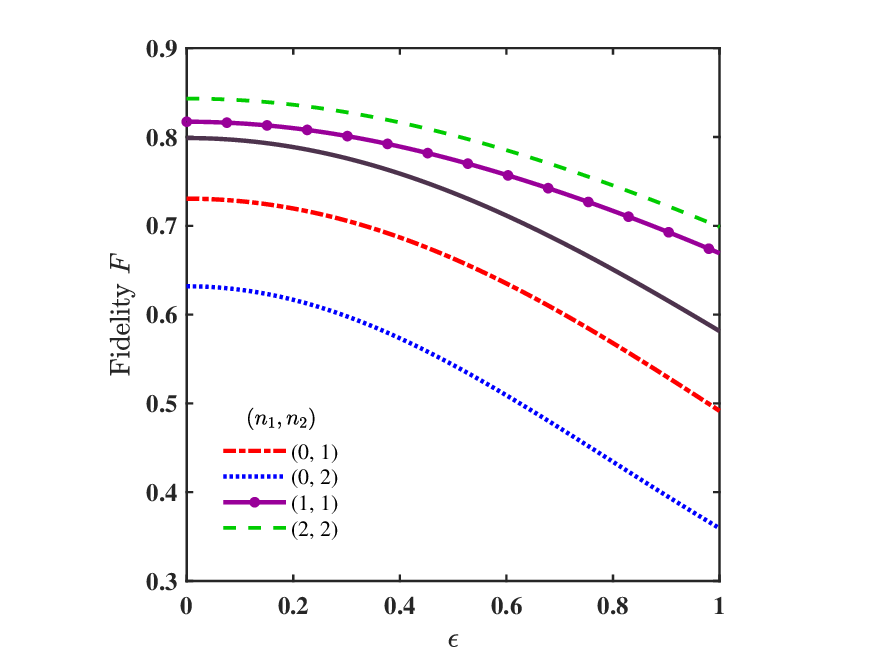}
	\caption{Teleportation fidelity $F$ as a function of the input squeezing parameter $\epsilon$ for different photon-subtraction configurations $(n_1,n_2)=(0,1)$, $(1,1)$, $(0,2)$, and $(2,2)$ of the PS-TMSFS resource. The resource state is prepared with $(m_1,m_2)=(1,1)$, beam-splitter transmissivities $T_1=T_2=0.9$, and the resource squeezing parameter is chosen as $\xi=0.8$. The solid black line corresponds to the Gaussian TMSV resource.
	}
	\label{Epi_seq}
\end{figure}
	Fig.~\ref{Epi_seq} shows the teleportation fidelity as a function of the input squeezing parameter $\varepsilon$ for different photon-subtraction configurations of the PS-TMSFS resource.  As the input squeezing increases, the teleportation fidelity decreases monotonically for all the considered resources. This behavior indicates that the highly squeezed input states are more challenging to teleport, irrespective of the resources employed. For all the investigated resources, the symmetric photon-subtraction configurations $(1,1)$ and $(2,2)$ again provide the highest teleportation fidelities throughout the entire range of $\varepsilon$ with the $(2,2)$ configuration consistently performing the best. The Gaussian TMSV resource yields moderate fidelity, remaining below the symmetric configurations while outperforming both the asymmetric photon-subtraction schemes. 
	
	Altogether, the results for coherent and squeezed vacuum input states demonstrate that the symmetric photon-subtraction consistently provides the maximum gain in the teleportation fidelity of Gaussian states. In particular, the balanced $(2,2)$ configuration emerges as the most effective PS-TMSFS resource over a broad range of squeezing parameters and beam-splitter transmissivities. After investigating the performance of the PS-TMSFS resource for the Gaussian input states, we now extend our analysis to the non-Gaussian inputs.

	\subsection{Teleportation of Non-Gaussian Input States}
	The analysis of the proposed PS-TMSFS channel is not restricted to Gaussian input states only. To further investigate the efficacy of the PS-TMSFS, we now focus on the teleportation of well-known non-Gaussian input states. As non-Gaussian states possess stronger quantum features, they provide a more rigorous assessment for the performance of the teleportation protocol. Since non-Gaussian states are more sensitive to losses introduced during the photon-subtraction process, a high beam-splitter transmissivity ($T=0.99$) is chosen to minimize these losses and to assess the performance of the PS-TMSFS resource under near-ideal subtraction conditions. In this subsection, we investigate the teleportation of two particular non-Gaussian input states, namely the Fock state and the Schr\"odinger cat state, using the PS-TMSFS resource.
	\subsubsection{Fock State}
	We first evaluate the teleportation fidelity for a single-mode Fock state $\ket{n}$ using the PS-TMSFS as the entangled resource. By substituting the Wigner characteristic functions of the Fock state given in Eq.~\eqref{Lag}, and the PS-TMSFS resource given in Eq.~\eqref{Normalise}, into Eq.~\eqref{F_B}, we obtain the teleportation fidelity as
	\begin{align}
		F &=	\frac{\hat{F}}{a_0}
		\sum_{k=0}^{n}
		\sum_{l=0}^{n}
		\frac{
			\binom{n}{k}
			\binom{n}{l}
		}
		{k!\,l!}
		\frac{\partial^{k+l}}
		{\partial z^{k+l}}
		\left[
		\frac{1}{z}
		\exp\!\left(
		u^{T} M_4 u
		\right)
		\right],
		\label{FFS}
	\end{align}
	where $\hat{F}$ denotes the differential operator and the quantity $z$ is defined in Eq.~\eqref{B-z} of Appendix~\ref{App_B}. Likewise the Gaussian input states, the fidelity depends on the squeezing parameter, the beam-splitter transmissivities, and the photon-subtraction configurations of the resource state. 
	In the following, we inspect how the squeezing parameter and the beam-splitter transmissivity affect the teleportation fidelity for different photon-subtraction configurations.
	\begin{figure}[H]
		\centering
		\includegraphics[trim=0pt 0pt 0pt 0pt,clip,width=0.5\textwidth]{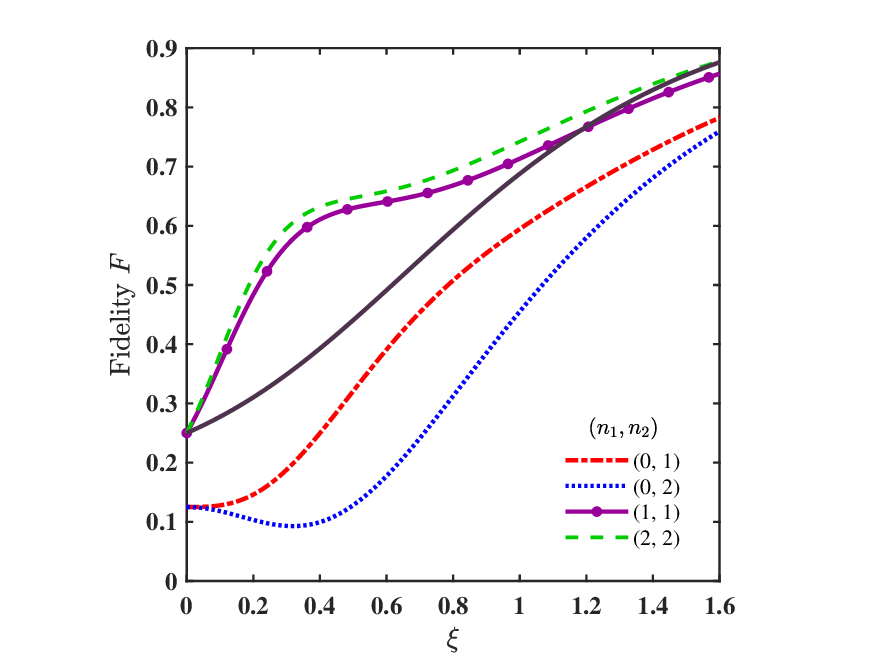}
		\caption{Teleportation fidelity $F$ for input Fock state $|1\rangle$ as a function of the squeezing parameter $\xi$ for different photon-subtraction configurations $(n_1,n_2)=(0,1)$, $(0,2)$, $(1,1)$ and $(2,2)$. The PS-TMSFS is prepared with $(m_1,m_2)=(1,1)$ and beam-splitter transmissivities $T_1=T_2=0.99$. The solid black line represents the teleportation fidelity for the input Fock state using the TMSV resource.}
		\label{R_Fock}
	\end{figure}
	Fig.~\ref{R_Fock} shows the teleportation fidelity for the input Fock state $|1\rangle$ as a function of the squeezing parameter $\xi$ using the PS-TMSFS resource. Similar to the input Gaussian scenarios, the symmetric photon-subtraction configurations $(1,1)$ and $(2,2)$ exhibit the higher teleportation fidelities over the entire squeezing range. Starting from nearly identical fidelities at $\xi=0$, both increase rapidly with the squeezing parameter $\xi$. The $(2,2)$ configuration remains the best-performing resource throughout the considered range, while the $(1,1)$ configuration closely follows it and differs slightly at large squeezing. The Gaussian TMSV resource, represented by the solid black curve, also increases  monotonically with the squeezing parameter. It yields lower fidelity than the symmetric configurations in the low- and intermediate-squeezing regimes but crosses $(1,1)$ at approximately $\xi\simeq1.2$, beyond which the TMSV resource exhibits improved fidelity. The $(2,2)$ configuration always outperforms the Gaussian TMSV resource and reaches the maximum fidelity value of $F=0.88$ at $\xi=1.6$. The asymmetric photon-subtraction configurations exhibit comparatively lower fidelities. Although the $(0,1)$ configuration increases steadily with squeezing, it remains below the Gaussian TMSV and both symmetric photon-subtraction schemes throughout the considered range. The fidelity of $(0,2)$ configuration exhibits a slight decrease at low-squeezing before increasing rapidly with $\xi$. Despite the improvement at larger squeezing values, its fidelity never exceeds those of the symmetric configurations. These results demonstrate that the balanced photon-subtraction provides a clear advantage for teleporting Fock states with the $(2,2)$ resource delivering the best overall performance over the entire squeezing regime.
	
	Next to the dependence of the teleportation fidelity on the squeezing parameter, we investigate the influence of the beam-splitter transmissivity employed in the photon-subtraction process of the PS-TMSFS resource.
	\begin{figure}[H]
		\centering
		\includegraphics[trim=0pt 0pt 0pt 0pt,clip,width=0.5\textwidth]{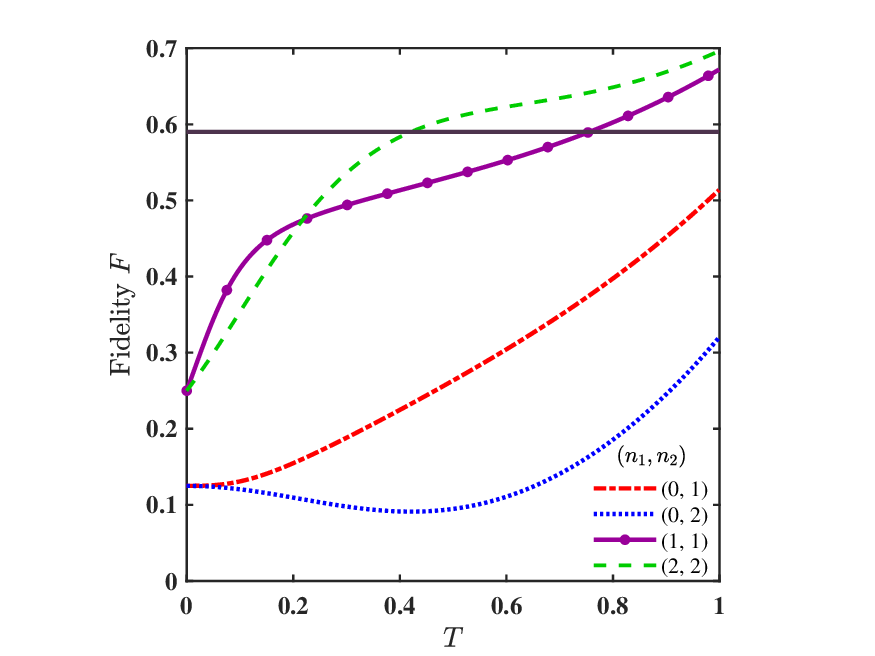}
		\caption{Teleportation fidelity $F$ for the input single-photon Fock state $\ket{1}$ as a function of the beam-splitter transmissivity $T$ ($T_1=T_2=T$) for different photon-subtraction numbers $(n_1,n_2)$. The PS-TMSFS is prepared with $(m_1,m_2)=(1,1)$ and the squeezing parameter is fixed at $\xi=0.8$. The solid black line denotes the fidelity of the Gaussian TMSV resource.}
		\label{T_Fock}
	\end{figure}
	Fig.~\ref{T_Fock} presents the teleportation fidelity for the input single-photon Fock state as a function of the beam-splitter transmissivity $T$ employed in the photon-subtraction process. The Gaussian TMSV resource, represented by the solid black line, is independent of the beam-splitter transmissivity and provides an upper bound for all the configurations at the low transmissivity values. For all the photon-subtraction configurations, the symmetric $(1,1)$ scheme initially provides the highest fidelity. As the beam-splitter transmissivity increases, the $(2,2)$ configuration exhibits a more pronounced enhancement in the teleportation fidelity and becomes the most effective photon-subtracted resource over the remaining transmissivity range. As the transmissivity increases further, the $(2,2)$ and $(1,1)$ configurations surpass the Gaussian TMSV resource. The asymmetric photon-subtraction schemes exhibit considerably lower fidelities throughout the considered range. The $(0,1)$ configuration increases monotonically with transmissivity but remains well below the symmetric schemes, barely exceeding the classical limit, whereas the $(0,2)$ configuration fails to even exceed the classical limit throughout the parameter range.
	These results demonstrate that the symmetric photon-subtraction, particularly the $(2,2)$ configuration, provides the maximum enhancement in the moderate and high-transmissivity regimes.
	\subsubsection{Schr\"odinger Cat State}
	We now consider the teleportation of a single-mode Schr\"odinger cat state, defined as a coherent superposition of two coherent states with opposite amplitudes $\ket{\pm{\gamma}}$ [see Eq.~\eqref{Cat_state}], using the PS-TMSFS as the entangled resource. By substituting the Wigner characteristic functions of the Schr\"odinger cat state $\chi_{\pm}$ [Eq.~\eqref{WCF_CAT}] and the PS-TMSFS into Eq.~\eqref{F_B}, the teleportation fidelity is obtained as 
	\begin{align}
		F
		&=
		\frac{\hat{F}}
		{d_0
			\left(
			1\pm e^{-2|\gamma|^{2}}
			\right)^{2}}
		\Biggl\{
		\exp\left(
		u^{T}M_4u
		\right)
		\nonumber\\
		&\quad\times
		\Bigg(
		\frac{1+e^{-4|\gamma|^{2}}}{2}
		+
		\frac{1}{2}
		\exp\left[
		-\frac{4a_0|\gamma|^{2}}{d_0}
		\right]
		\cosh\left[
		\frac{2(ipR+qS)}{d_0}
		\right]
		\nonumber\\
		&\qquad+
		\frac{e^{-4|\gamma|^{2}}}{2}
		\exp\left[
		\frac{4a_0|\gamma|^{2}}{d_0}
		\right]
		\cosh\left[
		\frac{2(qR+ipS)}{d_0}
		\right]
		\nonumber\\
		&\qquad\pm
		e^{-2|\gamma|^{2}}
		\cosh\left[
		\frac{qR+ipS}{d_0}
		\right]
		\cosh\left[
		\frac{qS+ipR}{d_0}
		\right]
		\Bigg)
		\Biggr\},
		\label{FCAT}
	\end{align}
	where $\hat{F}$ denotes the differential operator and $d_0=2\beta(\beta  +\alpha t_1t_2)$. Here $\gamma=(q+ip)/\sqrt{2}$ denotes the amplitude of the coherent state components of the Schr\"odinger cat state, while $R$ and $S$ are defined in Eq.~\eqref{R_S}. The effects of the squeezing parameter and the beam-splitter transmissivities are discussed below for different photon-subtraction configurations.
	\begin{figure}[H]
		\centering
		\includegraphics[trim=0pt 0pt 0pt 0pt,clip,width=0.5\textwidth]{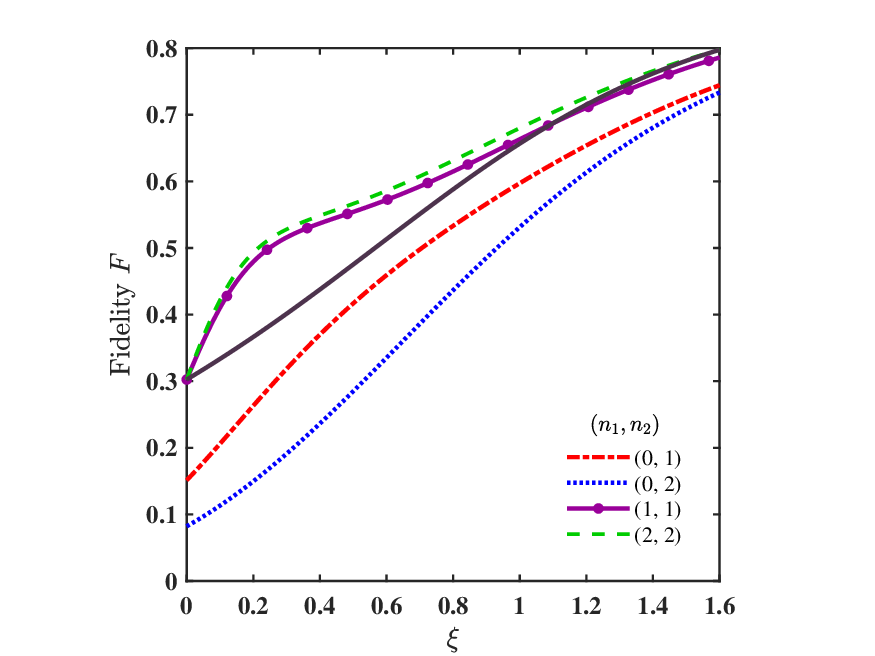}
		\caption{Teleportation fidelity $F$ for the input even Schr\"odinger cat state $\ket{\psi_{+}}$ as a function of the squeezing parameter $\xi$ for different photon-subtraction configurations $(n_1,n_2)$. The initial Fock state is fixed at $(m_1,m_2)=(1,1)$ with the coherent state amplitude $\gamma=1$ for the even cat state, and the beam-splitter transmissivities are set to $T_1=T_2=0.99$. The solid black line represents the teleportation fidelity for the input even cat state using the TMSV resource.}
		\label{R_cat}
	\end{figure}
	Fig.~\ref{R_cat} presents the teleportation fidelity for the input even Schr\"odinger cat state as a function of the squeezing parameter $\xi$ for different photon-subtraction configurations of the PS-TMSFS resource. Similar to the Fock state input, the symmetric photon-subtraction configurations $(1,1)$ and $(2,2)$ provide the higher teleportation fidelities throughout the entire squeezing range. Initially attaining the same fidelity values, both the configurations increase rapidly with the squeezing parameter $\xi$. The $(2,2)$ configuration consistently provides the highest fidelity, while the $(1,1)$ configuration closely tracks it throughout the considered range. The asymmetric photon-subtraction configurations $(0,1)$ and $(0,2)$ yield comparatively lower fidelities, although both increase monotonically with the squeezing parameter. The $(0,1)$ configuration outperforms the $(0,2)$ configuration over the entire squeezing range, whereas the Gaussian TMSV resource exhibits a smooth monotonic increase, surpassing the symmetric $(1,1)$ configuration at approximately $\xi\simeq1$. As the squeezing parameter increases, the difference among all the considered resources gradually decreases, and their fidelities approach nearly the same value ($\approx 0.8$) in the large-squeezing regime. This result indicates that the symmetric photon-subtraction provides the maximum enhancement in the low and intermediate-squeezing regimes, with the $(2,2)$ configuration offering the best overall performance. However, this advantage gradually diminishes as $\xi$ increases and the $(2,2)$ configuration becomes nearly indistinguishable with the Gaussian TMSV resource in terms of teleportation fidelity.
	\begin{figure}[H]
		\centering
		\includegraphics[trim=0pt 0pt 0pt 0pt,clip,width=0.5\textwidth]{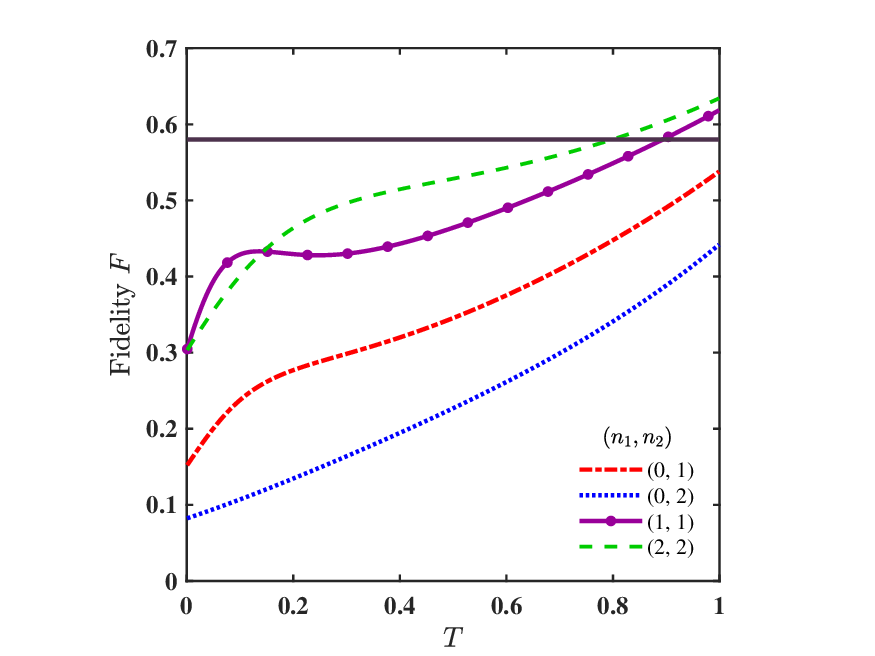}
		\caption{Teleportation fidelity $F$ for the input even Schr\"odinger cat state as a function of the beam-splitter transmissivity $T$ ($T_1=T_2=T$) for different photon-subtraction configurations $(n_1,n_2)$ of the PS-TMSFS resource. The underlying Fock state is prepared at $(m_1,m_2)=(1,1)$ with the coherent state amplitude $\gamma=1$ for the even cat state, and the squeezing parameter is set to $\xi=0.8$. The solid black line shows the fidelity for the Gaussian TMSV resource.}
		\label{T_cat}
	\end{figure}
	
	Fig.~\ref{T_cat} illustrates the dependence of the teleportation fidelity for the input even Schr\"odinger cat state on the beam-splitter transmissivity employed for preparing the PS-TMSFS resource. The dependence on transmissivity is analogous to that obtained for the input single-photon Fock state. The Gaussian TMSV resource provides the highest fidelity in the low and intermediate-transmissivity region. Among the condidered photon-subtraction patterns, the symmetric $(1,1)$ configuration performs better at low transmissivity and is subsequently surpassed by the $(2,2)$ configuration as $T$ increases. The asymmetric $(0,1)$ and $(0,2)$ configurations yield considerably lower fidelities throughout the considered range with the $(0,2)$ scheme remaining below the classical limit for all the transmissivity values. Although, the symmetric photon-subtraction provides a clear advancement over the corresponding asymmetric configurations and eventually surpasses the Gaussian TMSV resource, none of the considered resources exceed the no-cloning limit \cite{Cerf2000, Grosshans2001} for the selected value of the squeezing parameter. This indicates that the degree of entanglement is insufficient to support the high-fidelity teleportation of an even Schr\"odinger cat state. Therefore, a larger squeezing parameter is required to achieve the successful quantum teleportation, highlighting the demand for a better resource for the teleportation of a substantially non-Gaussian input state.
	
	Overall, the results show that the performance of the PS-TMSFS resource depends strongly on the squeezing parameter, beam-splitter transmissivity, photon-subtraction configuration, and the nature of the input state. The symmetric photon-subtraction consistently enhances the fidelity over the Gaussian TMSV resource in the low-squeezing regime, an advantage that diminishes as squeezing increases, whereas the asymmetric configurations remain less effective throughout the range. With respect to the transmissivity, the TMSV resource performs best at low $T$ values, while the symmetric photon-subtraction becomes advantageous for higher values of $T$ and the $(2,2)$ configuration performs the best among the considered PS-TMSFS resources. Although, a significant improvement is achieved for the Gaussian input states, the teleportation of the non-Gaussian inputs remains more challenging. For these states, higher squeezing and beam-splitter transmissivity are generally required to achieve higher teleportation fidelities, highlighting that they are more sensitive to losses and imperfections in the resource preparation. 	\\		
	The dependence of the teleportation fidelity on the beam-splitter transmittance $T$ can be understood from the attenuation introduced by the photon-subtraction process. The passage of the two-mode squeezed Fock state through the photon-subtraction beam splitters effectively introduces a noiseless attenuation, which reduces the two-mode squeezing factor $\tanh\xi$ by a factor of $T$~\cite{Miff2012, Nunn2022}. For $T\rightarrow1$, this attenuation becomes weak, and the conditional operation approaches the ideal photon-subtraction regime. As $T$ decreases, the attenuation becomes increasingly significant and suppresses the two-mode squeezing and, consequently, the correlations required for teleportation. Thus, the fidelity is determined by the balance between the non-Gaussian enhancement due to photon subtraction and the suppression of two-mode squeezing caused by attenuation. A sufficiently large transmittance is therefore required for the non-Gaussian advantage to overcome the detrimental effect of attenuation and surpass the conventional Gaussian TMSV protocol.
	The findings indicate that the appropriately engineered symmetric photon-subtraction offers the maximum benefit for CV quantum teleportation, especially in the experimentally realizable low-squeezing regime.
		
\section{Conclusion}
\label{conclusion}

In this paper, we investigate the applicability of the photon-subtracted two-mode squeezed Fock state (PS-TMSFS) as an entangled resource in continuous-variable quantum teleportation protocol. The success probability corresponding to the conditional photon-subtraction process is evaluated and the Wigner characteristic function of the resulting non-Gaussian resource state is derived analytically. Using these, the teleportation fidelities for Gaussian and non-Gaussian input states are determined and analyzed for a variety of squeezing parameters, beam-splitter transmissivities, and different photon-subtraction configurations.	Our results show that the teleportation fidelity using the PS-TMSFS channel exhibits a strong dependence on the squeezing parameter, beam-splitter transmissivity, photon-subtraction configuration, and the nature of the input state. Despite its lower success probability compared with the asymmetric schemes, the symmetric photon-subtraction consistently outperforms the asymmetric configurations with the $(2,2)$ resource achieving the highest fidelity across most of the parameter range studied. For teleporting a Gaussian input, the PS-TMSFS offers a clear advantage over the Gaussian TMSV resource across a wide range of the squeezing and transmissivity parameters. For non-Gaussian input states, the symmetric $(2,2)$ photon-subtraction remains the best-performing PS-TMSFS configuration but the Gaussian TMSV resource continues to be a viable alternative and even surpasses the PS-TMSFS in certain regimes. For both the inputs, the fidelity differences among the photon-subtraction configurations diminish with the increasing squeezing, highlighting that the advantage of photon subtraction is most pronounced in the experimentally realizable low-squeezing regime.
	
	The present findings establish the PS-TMSFS as a promising non-Gaussian entangled resource for CV quantum teleportation and provide practical insights for selecting an appropriate photon-subtraction configuration based on the input state class and the achievable squeezing. In general, the results indicate that the balanced photon-subtraction can effectively enhance CV quantum teleportation, particularly for Gaussian input states. For non-Gaussian input states, achieving higher fidelity generally requires stronger squeezing and higher beam-splitter transmissivity, motivating future work on alternative photon-subtraction and photon-addition schemes and their associated trade-off between the fidelity enhancement and the success probability.

\appendix
\section{Phase-space description of the continuous-variable system}
\label{App_A}

In the present work, we adopt the phase-space formulation as it provides an appropriate framework for characterizing the quantum states and evaluating the considered teleportation protocol. In this appendix, we summarize the phase-space formalism of continuous-variable quantum systems that includes the basic definitions of the quadrature operators, the descriptions of the quantum states in terms of the quasiprobability distributions and the associated Wigner characteristic functions.

For an $n$-mode bosonic system, we define the vector of canonical quadrature operators as
\begin{equation}
	\hat{\boldsymbol{\eta}} 
	=
	(\hat{q}_1,\hat{p}_1,\dots,\hat{q}_n,\hat{p}_n)^T,
\end{equation}
which satisfies the commutation relations
\begin{equation}
	[\hat{\eta}_i,\hat{\eta}_j]
	=
	i\Omega_{ij},\qquad i, j=1, 2, \ldots 2n,
\end{equation}
where the symplectic matrix $\Omega$ is given by
\begin{equation}
	\Omega
	=
	\bigoplus_{k=1}^{n}\omega,\qquad \omega=
	\begin{pmatrix}
		0 & 1\\
		-1 & 0
	\end{pmatrix}.
\end{equation}
Equivalenly, the state of the system can be described in terms of the bosonic annihilation and creation operators as
\begin{equation}
	\hat{a}_i
	=
	\frac{1}{\sqrt{2}}
	(\hat{q}_i+i\hat{p}_i),
	\qquad
	\hat{a}_i^\dagger
	=
	\frac{1}{\sqrt{2}}
	(\hat{q}_i-i\hat{p}_i).
\end{equation}
For instance, the Fock state $\ket{n_i}$ can be generated by repeated application of the creation operator on the vacuum state as
\begin{equation}
	\ket{n_i}
	=
	\frac{(\hat{a}_i^\dagger)^{n_i}}{\sqrt{n_i!}}
	\ket{0},
\end{equation}
and satisfies
\begin{equation}
	\hat{n_i}	\ket{n_i}
	=\hat{a}_i^\dagger \hat{a}_i \ket{n_i}
	=
	n_i \ket{n_i}.
\end{equation}
Here $\hat{n}_i$ is the photon number operator acts on mode $i$. For a multi-mode system, the Hilbert space is constructed as the tensor product of single-mode Fock spaces. In particular, a two-mode system in Fock basis can be written as
\begin{equation}
	\ket{m_1,m_2}
	=
	\ket{m_1}\otimes\ket{m_2}.
\end{equation}

A convenient way to describe an $n\text{-mode}$ quantum system with a density operator $\hat{\rho}$ in phase-space is the Wigner characteristic function defined as
\begin{equation}
	\chi(\Lambda)
	=
	\mathrm{Tr}
	\left[
	\hat{\rho}\,
	e^{-i\Lambda^T\Omega\hat{\boldsymbol{\eta}}}
	\right],
\end{equation}
where $\Lambda=(\Lambda_1, \Lambda_2, \ldots, \Lambda_n)^T \in\mathbb{R}^{2n}$ denotes the phase-space variable. The first-order moment for an $n$-mode system is defined as
\begin{equation}
	\mathrm{d}
	=
	\langle\hat{\eta}\rangle= \mathrm{Tr}[{\hat{\rho}\hat{\eta}}],
\end{equation}
while the second-order moment or covariance matrix is given by
\begin{equation}
	V_{ij}
	=
	\frac{1}{2}
	\left\langle
	\{
	\Delta\hat{\eta}_i,
	\Delta\hat{\eta}_j
	\}
	\right\rangle,
\end{equation}
with $\Delta\hat{\eta}_i=\hat{\eta}_i-\langle\hat{\eta}_i\rangle$. 
In particular, the Wigner characteristic function associated with the single-mode Fock state $\ket{n}$ is
\begin{equation}
	\chi_{|n\rangle}(\tau,\sigma)
	=
	\exp\left[
	-\frac{\tau^2}{4}
	-\frac{\sigma^2}{4}
	\right]
	L_n
	\left(
	\frac{\tau^2+\sigma^2}{2}
	\right),
	\label{Lag}
\end{equation}
where $L_n(x)$ denotes the Laguerre polynomial of order $n$. For convenience, Eq.~(\ref{Lag}) may also be expressed in terms of the exponential generating function as 
\begin{equation}
	\chi_{|n\rangle}(\tau,\sigma)
	=
	\exp\left[
	-\frac{\tau^2}{4}
	-\frac{\sigma^2}{4}
	\right]
	\hat{F}
	e^{2st+s(\tau+i\sigma)-t(\tau-i\sigma)},
\end{equation}
where
\begin{equation}
	\hat{F}
	=
	\frac{1}{2^n n!}
	\frac{\partial^n}{\partial s^n}
	\frac{\partial^n}{\partial t^n}\{\bullet\}
	\Bigg|_{s=t=0}.
\end{equation}
A Gaussian state is characterized by a Wigner function that displays the Gaussian form in phase-space. Such states are completely determined by their first and second-order statistical moments, namely the displacement vector and the covariance matrix. So for a Gaussian state described by the displacement vector $\mathrm{d}$ and the covariance matrix $V$, the corresponding Wigner characteristic function can be written as

\begin{equation}
	\chi(\Lambda)
	=
	\exp\!\left[
	-\frac{1}{2}\Lambda^{T}(\Omega V \Omega^T) \Lambda
	-i(\Omega\mathrm{d})^T \Lambda
	\right]. 
	\label{WCF_GS}
\end{equation}
The Gaussian operations related to the present study are represented by the symplectic transformations in phase-space. In phase-space, the coherent state corresponds to a trivial symplectic transformation as
\begin{equation}
	S = I,
\end{equation}
and the associated unitary operator is
\begin{equation}
	U(D(\alpha))
	=
	\exp\left[
	\alpha \hat{a}^\dagger - \alpha^* \hat{a}
	\right].
\end{equation}
For a single-mode coherent state with displacement vector $\mathrm{d}=(d_q,\,d_p)^T$, the Wigner characteristic function \eqref{WCF_GS} reduces to
\begin{equation}
	\chi_{\mathrm{coh}}(\Lambda)
	=
	\exp\!\left[
	-\frac{1}{4}(\tau^2+\sigma^2)
	-i(\tau d_p-\sigma d_q)
	\right].
	\label{WCF_CS}
\end{equation}
Unlike a coherent state, a squeezed state arises from a non-trivial symplectic transformation that redistributes the quantum uncertainties between the conjugate quadratures while maintaining the canonical commutation relations. A single-mode squeezing operation is represented by the symplectic matrix
\begin{equation}
	S(r)
	=
	\begin{pmatrix}
		e^{-r} & 0\\
		0 & e^{r}
	\end{pmatrix},
\end{equation}
where \(r\) is the squeezing parameter. This transformation is represented by the unitary operator as
\begin{equation}
	U(S(r))
	=
	\exp\left[
	\frac{r}{2}
	(\hat{a}^2-\hat{a}^{\dagger 2})
	\right].
\end{equation}
The Wigner characteristic function \eqref{WCF_GS} of a single-mode squeezed vacuum state generated by the above squeezing transformation is given by
\begin{equation}
	\chi_{\mathrm{squ}}(\Lambda)
	=
	\exp\!\left[
	-\frac{1}{4}
	\left(
	\tau^2 e^{2r}
	+
	\sigma^2 e^{-2r}
	\right)
	\right].
	\label{WCF_SS}
\end{equation}
The two-mode squeezing symplectic transformation that entangles two individual modes is
\begin{equation}
	S_2(r)
	=
	\begin{pmatrix}
		\cosh r\, \mathbb{I}_2 & \sinh r\, \mathbb{Z}\\\\
		\sinh r\, \mathbb{Z} & \cosh r\, \mathbb{I}_2
	\end{pmatrix},
\end{equation}
where $\mathbb{Z}=\mathrm{diag}(1,-1)$. Its unitary representation is given by
\begin{equation}
	S_2(r)
	=
	\exp\left[
	r
	(
	\hat{a}_i^\dagger\hat{a}_j^\dagger
	-
	\hat{a}_i\hat{a}_j
	)
	\right],
	\label{TMSO}
\end{equation}
	which can be written in disentangled form as 
	\begin{equation} 
		S_2(r) = \exp\!\left[ \tanh (r)\,\hat{a}_1^{\dagger}\hat{a}_2^{\dagger} \right] (\sech (r))^{\hat{n}_1+\hat{n}_2+1} \exp\!\left[ -\tanh (r)\,\hat{a}_1\hat{a}_2 \right]. \label{TMSV_disentangle} \end{equation}

Further, the beam-splitter interaction between the two modes is represented by
\begin{equation}
	B_{ij}(T)
	=
	\begin{pmatrix}
		\sqrt{T}\,\mathbb{I}_2 & \sqrt{1-T}\,\mathbb{I}_2\\\\
		-\sqrt{1-T}\,\mathbb{I}_2 & \sqrt{T}\,\mathbb{I}_2
	\end{pmatrix},
\end{equation}
where $T$ denotes the beam-splitter transmissivity. The corresponding unitary operator is given by
\begin{equation}
	U(B_{ij}(T))
	=
	\exp\left[
	\sqrt{T}
	(
	\hat{a}_i^\dagger\hat{a}_j
	-
	\hat{a}_j^\dagger\hat{a}_i
	)
	\right].
\end{equation}
Under a homogeneous symplectic transformation $S$, the density operator transforms as
\begin{equation}
	\hat{\rho}
	\rightarrow
	U(S)\hat{\rho} U^\dagger(S),
\end{equation}
which induces the following transformations in phase space
\begin{equation}
	\mathrm{d}\rightarrow S\mathrm{d},
	\qquad
	V\rightarrow SVS^T,
	\qquad
	\chi(\Lambda)\rightarrow\chi(S^{-1}\Lambda).
\end{equation}

The even ($+$) and odd ($-$) Schr\"odinger cat states are defined as a coherent superposition of two coherent states with opposite amplitudes as
	\begin{equation}
		\ket{\psi_{\pm}}
		=
		\mathcal{N}_{\pm}
		\left(
		\ket{\gamma}
		\pm
		\ket{-\gamma}
		\right),
		\label{Cat_state}
	\end{equation}
	where the normalization constant is
	\begin{equation}
		\mathcal{N}_{\pm}
		=
		\frac{1}
		{\sqrt{2\left(1\pm e^{-2|\gamma|^{2}}\right)}}.
	\end{equation}
	The corresponding Wigner characteristic function is given by
	\begin{equation}
		\chi_{\pm}(\alpha)
		=
		\frac{
			e^{-|\alpha|^{2}/2}
		}
		{1\pm e^{-2|\gamma|^{2}}}
		\left[
		\cosh
		\left(
		\alpha\gamma^{*}
		-
		\alpha^{*}\gamma
		\right)
		\pm
		e^{-2|\gamma|^{2}}
		\cosh
		\left(
		\alpha\gamma^{*}
		+
		\alpha^{*}\gamma
		\right)
		\right].
		\label{WCF_CAT}
\end{equation}

\section{Matrices and coefficients appearing in the Wigner characteristic functions, success probability and fidelity of QT using PS-TMSFS}
\label{App_B}

\subsection*{1. Wigner characteristic function and success probability of PS-TMSFS}
The matrices appearing in the Wigner characteristic function of the PS-TMSFS [see Eq.~\eqref{eq11}] are given explicitly
%
\begin{align}
	M_1 &= 	-\frac{1}{4{a}_0}	\begin{pmatrix}
		a_1 & 0 & -a_2 &0 \\
		0 & a_1 & 0 & a_2 \\
		-a_2 & 0 & a_1 & 0 \\
		0& a_2 & 0 & a_1
	\end{pmatrix}  \label{M_1}\\[0.2cm]
	\text{with} ~~~~~~ & {a}_0 =  \beta^2 -\alpha^2 t_1^2 t_2^2, \nonumber \\ & {a}_1 =  \beta^2 +\alpha^2 t_1^2 t_2^2, \\  &{a}_2 =  2\alpha\beta t_1 t_2.\nonumber\\\nonumber \\
	\text{Here} ~~~~~~~ & t_1=\sqrt{T_1},\,t_2=\sqrt{T_2}, \nonumber \\ &r_1^2+t_1^2=1,\,\,r_2^2+t_2^2=1,  \\ & \alpha=\sinh{\xi}, \nonumber \\
	& \beta=\cosh{\xi}\nonumber, 
\end{align}
\begin{align}
	M_2 = \frac{1}{a_0}
	\begin{pmatrix}
		b_1 & ib_1 & b_2 & -ib_2 \\
		-b_1 & ib_1 & -b_2 & -ib_2 \\
		b_3 & -ib_3 & b_4 & ib_4 \\
		-b_3 &- ib_3 &- b_4 & ib_4 \\
		b_5 & ib_5 & b_6 & -ib_6 \\
		-b_5 & ib_5 & -b_6 & -ib_6 \\
		b_7 & -ib_7 & b_8 & ib_8 \\
		-b_7 & -ib_7 & -b_8 & ib_8
	\end{pmatrix},  \label{M_2} \\[0.2cm]
	\begin{aligned}
		& b_1 = \beta t_1, \quad && b_5 = \alpha^2 r_1 t_1 t_2^2, \\
		& b_2 = -\alpha  t_1^2 t_2, \quad && b_6 = -\alpha \beta r_1 t_2, \\
		& b_3 = -\alpha t_1 t_2^2, \quad && b_7 = -\alpha \beta r_2 t_1, \\
		& b_4 = \beta t_2, \quad && b_8 = \alpha^2 r_2 t_1^2 t_2
	\end{aligned} 
\end{align}
\begin{align}
	M_3 = \frac{1}{a_0}
	\begin{pmatrix}
		0 & c_1 & c_2 & 0 &  0 &c_3 & c_4 & 0  \\
		c_1 & 0 & 0 & c_2 & c_3 & 0 &0 & c_4 \\
		c_2 & 0 & 0 & c_5 &  c_6 & 0 & 0 & c_7 \\
		0 & c_2 &  c_5 & 0 & 0 & c_6 & c_7 & 0 \\
		0 & c_3 & c_6 & 0 &  0 &c_8 & c_9 & 0  \\
		c_3 & 0 & 0 & c_6 & c_8 & 0 &0 & c_9 \\
		c_4 & 0 & 0 & c_7 &  c_9 & 0 & 0 & c_{10} \\
		0 & c_4 &  c_7 & 0 & 0 & c_9 & c_{10} & 0 
	\end{pmatrix}, \label{M_3}
	\\[0.2cm]
	\begin{aligned}
		& c_1 = t_1^2, \quad && c_6 = \alpha  r_1  t_2^2, \\
		& c_2 = -\alpha \beta(1-t_1^2 t_2^2), \quad && c_7 = \beta r_2 , \\
		& c_3 = r_1 \beta , \quad && c_8 = \alpha^2 r_1^2 t_2^2, \\
		& c_4 = \alpha r_2 t_1^2, \quad && c_9 = \alpha \beta r_1 r_2, \\
		& c_5 = t_2^2, \quad && c_{10} = \alpha^2 t_1^2 r_2^2 	\end{aligned}  
\end{align}
\subsection*{2. Fidelity for input Gaussian states using PS-TMSFS}
The explicit form of the matrix $M_4$ appearing in the teleportation
fidelity of the input coherent state [in Eq.~\eqref{FCS}] is given by
\begin{align}
	M_4 = \frac{1}{d_0}
	\begin{pmatrix}
		0 & c_1 & d_1 & 0 & 0 & d_2 & c_4 & 0 \\
		c_1 & 0 & 0 & d_1 & d_2 & 0 & 0 & c_4 \\
		d_1 & 0 & 0 & c_5 & c_6 & 0 & 0 & d_3 \\
		0 & d_1 & c_5 & 0 & 0 & c_6 & d_3 & 0 \\
		0 & d_2 & c_6 & 0 & 0 & c_8 & d_4 & 0 \\
		d_2 & 0 & 0 & c_6 & c_8 & 0 & 0 & d_4 \\
		c_4 & 0 & 0 & d_3 & d_4 & 0 & 0 & c_{10} \\
		0 & c_4 & d_3 & 0 & 0 & d_4 & c_{10} & 0
	\end{pmatrix} \label{M_4}
\end{align}
where 
\begin{eqnarray}
	\begin{aligned}
		& d_0 = 2\beta (\beta - \alpha t_1t_2), \\ & d_1=\frac{ t_1 t_2 (\beta - \alpha t_1 t_2) + 2\beta c_2}
		{\beta+ \alpha t_1 t_2}, \\ & d_2 = r_1 (2\beta - \alpha t_1t_2), \\ & d_3 = r_2 (2\beta - \alpha t_1t_2), \\& d_4 = \alpha r_1 r_2 (2\beta - \alpha t_1t_2)
	\end{aligned}
\end{eqnarray}
The explicit form of the matrix $M_5$ appearing in the teleportation fidelity of the input squeezed state [see Eq.~\eqref{FSS}] is
\begin{align}
	M_5& = \frac{1}{e_0}
	\begin{pmatrix}
		e_1 & e_2 & e_3 & e_4 & e_5 & e_6 & e_7 & e_8 \\
		e_2 & e_1 & e_4 & e_3 & e_6 & e_5 & e_8 & e_7 \\
		e_3 & e_4 & e_9 & e_{10} & e_{11} & e_{12} & e_{13} & e_{14} \\
		e_4 & e_3 & e_{10} & e_9 & e_{12} & e_{11} & e_{14} & e_{13} \\
		e_5 & e_6 & e_{11} & e_{12} & e_{15} & e_{16} & e_{17} & e_{18} \\
		e_6 & e_5 & e_{12} & e_{11} & e_{16} & e_{15} & e_{18} & e_{17} \\
		e_7 & e_8 & e_{13} & e_{14} & e_{17} & e_{18} & e_{19} & e_{20} \\
		e_8 & e_7 & e_{14} & e_{13} & e_{18} & e_{17} & e_{20} & e_{19}
	\end{pmatrix} 
	\label{M_5} 
\end{align}
where
\begin{eqnarray}
	\begin{aligned}
		& e_0= 2 (a_1+\delta a_0) , \\& e_1= -\mu t_1^2,  \\
		& e_2= t_1^2\left(\frac{ a_0}{a_3} + \delta \right) , \\& e_3= \frac{1}{a_0}
		\left(
		\frac{t_1 t_2}{2}(a_3+\delta a_0)
		- \alpha\beta(1-t_1^2 t_2^2)e_0
		\right), \\
		& e_4= \mu t_1 t_2 , \\& e_5= -\mu t_1 \alpha r_1 t_2, \\
		& e_6= \frac{r_1}{a_0}
		\left(	(\alpha t_1 t_2 + 2 \beta)e_0
		-	\alpha t_1 t_2a_2	\right) , \\& e_7= \alpha r_2 t_1^2\left(\frac{ a_0}{a_3} + \delta \right), \\
		& e_8= -\mu \alpha r_2 t_1^2 , \\& e_9= -\mu t_2^2, \\
		& e_{10}= t_2^2\left(\frac{ a_0}{a_3} + \delta \right), \\& e_{11}= \alpha r_1 t_2^2\left(\frac{ a_0}{a_3} + \delta \right), \\
		& e_{12}= -\mu \alpha r_1 t_2^2 , \\ & e_{13}= \mu \alpha r_2 t_1 t_2, \\
		& e_{14}= \frac{r_2}{a_0}
		\left(	(\alpha t_1 t_2 + 2 \beta)e_0
		-	\alpha t_1 t_2a_2	\right) , \\& e_{15}= -\mu \alpha^2 r_1^2 t_2^2, \\
		& e_{16}= \alpha^2 r_1^2 t_2^2\left( \frac{ a_0}{a_3} + \delta \right) , \\& e_{17}= \frac{\alpha r_1 r_2}{a_0}
		\left(	(\alpha t_1 t_2 + 2 \beta)e_0
		-	\alpha t_1 t_2a_2	\right), \\
		& e_{18}= \mu \alpha r_1 r_2 t_1 t_2 , \\& e_{19}= - \mu \alpha^2 r_2^2 t_1^2, \\
		& e_{20}= \alpha^2 r_2^2 t_1^2 \left( \frac{ a_0}{a_3} + \delta \right).  \\\\
		\text{Here,}~~~~~~
		& \mu=\sinh{2\epsilon},\\
		&\delta=\cosh{2\epsilon}.
	\end{aligned}
\end{eqnarray} 

\subsection*{3. Fidelity for input non-Gaussian states using PS-TMSFS}
	
	The quantities $z$, $R$ and $S$ appeared in \eqref{FFS} and \eqref{FCAT} are
	\begin{align}
		z=\frac{a_1-a_2}{a_0}+1,
		\label{B-z}
	\end{align}
	and
	\begin{align}
		R&=u^T
		\begin{bmatrix}
			-(b_1+b_2)\\
			b_1+b_2\\
			-(b_3+b_4)\\
			b_3+b_4\\
			-(b_5+b_6)\\
			b_5+b_6\\
			-(b_7+b_8)\\
			b_7+b_8
		\end{bmatrix},
		\qquad
		S=u^T
		\begin{bmatrix}
			b_1+b_2\\
			b_1+b_2\\
			-(b_3+b_4)\\
			-(b_3+b_4)\\
			b_5+b_6\\
			b_5+b_6\\
			-(b_7+b_8)\\
			-(b_7+b_8)
		\end{bmatrix} \label{R_S}
	\end{align}
\section*{Acknowledgements}	
Ankita's work is supported by the University Grants Commission (UGC), Government of India (Award No.~231610110670). A.~C. acknowledges Haryana State Council for Science, Innovation and Technology for the support provided through the Project No. HSCSIT/R$\&$D/2025/2060. 
\bibliography{paper2_ref}
\end{document}